\documentclass{appolb}
\usepackage{graphicx}
\usepackage{amssymb,amscd,amsmath}
\usepackage{wrapfig}
\usepackage{feynmf}
\usepackage{appendix}
\usepackage{slashed}
\usepackage{indentfirst}
\usepackage{wrapfig}
\usepackage{appendix}

\newcommand{\absvec}[1]{\left| \mathbf{ #1 } \right|}
\begin{document}
% \eqsec  % uncomment this line to get equations numbered by (sec.num)
\title{Turbulence-Induced Instabilities in EP and QGP%
\thanks{Based on the talk presented by A. Leonidov at ISMD 2012, Kielce, Poland}%
% you can use '\\' to break lines
}
\author{Martin Kirakosyan${}^{(a)}$, Andrei Leonidov${}^{(a,b,c)}$
\address{(a) P.N. Lebedev Physical Institute, Leninsky pr. 53, 119991 Moscow, Russia \\
(b) Institute of Theoretical and Experimental Physics, Moscow, Russia \\
(c) Moscow Institute of Physics and Technology, Moscow, Russia}
\medskip
\medskip
\\
{Berndt M\"{u}ller
}
\address{Department of Physics, Duke University, Durham, NC 27708, USA}
}
\maketitle
\begin{abstract}
Polarization properties of turbulent stochastically inhomogeneous ultrarelativistic QED plasma are studied. It is shown that the sign of nonlinear
turbulent Landau damping corresponds to an instability of the spacelike modes and, for sufficiently large turbulent fields, to an actual
instability of a system.
\end{abstract}

\section{Introduction}

Working out a quantitative description of the properties of dense strongly interacting matter produced in ultrarelativistic heavy ion collisions
presents one of the most fascinating problems in high energy physics. The simplest (albeit not unique) way of putting the experimental data from
RHIC \cite{hydroRHIC} and LHC \cite{hydroLHC} into a coherent framework is to describe the essential physics of these collisions as a
hydrodynamical expansion of primordial quark-gluon matter that, after a short transient period, reaches sufficient level of local equilibration
allowing the usage of hydrodynamics. The features of the experimentally observed energy flow, in particular the presence of a strong elliptic
flow, suggest early equilibration of the initially produced matter and small shear viscosity of the expanding fluid, see e.g. the discussion in
\cite{H05} and \cite{MSW12} devoted to RHIC and LHC results respectively.

Can stylized features of primordial quark-gluon matter, in particular its anomalously low viscosity, be described within a weakly coupled theory,
i.e. as a plasma composed of quasiparticles with the quantum numbers of  quarks and gluons?
To address this question let us recall that extensive experimental studies of "ordinary"\;
electromagnetic plasma has demonstrated, see e.g. \cite{Tsit}, that it is practically never observed in the state of textbook thermal equilibrium.
Realistic description of the properties of experimentally observed QED plasma is possible only through taking into account the presence, in addition
to thermal excitations, of randomly excited fields. The resulting state was termed {\em turbulent plasma}. Collective properties of turbulent plasmas
are markedly different from those of the ordinary equilibrium plasmas. In particular, they are characterized by anomalously low shear viscosity and
conductivity, dominant effects of coherent nonlinear structures on transport properties.

Thus it is natural to consider turbulent QCD plasma as a natural candidate for describing the primordial quark-gluon matter in the weak coupling
regime. Calculation of shear viscosity of turbulent QGP performed in \cite{ABM} has indeed demonstrated that its shear viscosity is anomalously
small. In the present paper we focus on studying the leading turbulent contributions to polarization properties of turbulent relativistic plasma.
For simplicity we restrict our consideration to the Abelian case. The non-Abelian generalization is briefly described in Section~\ref{conc}.

\section{Turbulent polarization}

\subsection{Theoretical formalism}

A weakly turbulent plasma is described as perturbation of an equilibrated system of (quasi-)particles by weak turbulent fields $F^T_{\mu \nu}$. In the
collisionless Vlasov approximation, the plasma properties are defined by the following system of equations ($F^R_{\mu \nu}$ is a regular non-turbulent
field):
\begin{eqnarray}
&& p^{\mu}\left[ \partial_{\mu}- e q \left( F^R_{\mu \nu} + F^T_{\mu \nu} \right) \dfrac{\partial}{\partial p_{\nu}}\right]f(p,x,q)=0 \nonumber\\
&&\partial^{\mu}\left( F^R_{\mu \nu} + F^T_{\mu \nu} \right)= j_{\nu}(x) = e \sum_{q,s}\int dp\, p_{\nu}\, q\, f(p,x,q) .
\label{kinetic+maxw}
\end{eqnarray}
The stochastic ensemble of turbulent fields is assumed to be Gaussian and characterized by the following correlators:
\begin{equation}
\langle F_{\mu \nu}^{T}\rangle=0, \;\;\;\;\;  \langle F^{T \mu \nu}(x)F^{T \mu^{\prime} \nu^{\prime}}(y)\rangle=K^{\mu \nu \mu^{\prime} .
\nu^{\prime}}(x,y) \label{turbens}
\end{equation}
In the present study we use the following parametrization of the two-point correlator $K^{\mu \nu \mu^{\prime} \nu^{\prime}}(x,y) $ \cite{ABM}:
\begin{equation}
K^{\mu \nu \mu^{\prime} \nu^{\prime}}(x)=K_{0}^{\mu \nu \mu^{\prime} \nu^{\prime}}\exp\left[-\dfrac{t^{2}}{2\tau^{2}}-\dfrac{r^{2}}{2 a^{2}}
\right]
\end{equation}

Turbulent polarization arises as a (linear) response to a regular perturbation that depends on turbulent fields. It is fully characterized by the
polarization tensor $\Pi^{\mu \nu} (k)$ defined as a variational derivative of the averaged induced current $\langle   j^{\mu}( k \; \vert
F^R,F^T) \rangle_{F^T}$ over the regular gauge potential $A^R_{\nu}$:
\begin{eqnarray}
&&\Pi^{\mu \nu} (k) = \frac{ \delta \langle   j^{\mu}(k \vert F^R,F^T) \rangle_{F^T}}{\delta A^R_{\nu}} \label{PolarTensDef}  \\
&&\langle   j^{\mu}( k \; \vert F^R,F^T) \rangle_{F^T} = e \sum_{q,s} \int d P p_{\nu} q \langle \delta f (p,k,q \vert F^R,F^T) \rangle_{F^T}
\label{incur}
\end{eqnarray}
Let us rewrite the kinetic equation in (\ref{kinetic+maxw}) in the following condensed form
\begin{equation}
f=f^{eq} + G p^{\mu}F_{\mu\nu}\partial_{p}^{\mu}f \; , \;\;\;\; G \equiv \dfrac{e q}{\imath((pk)+\imath \epsilon)},
\end{equation}
where $f^{eq}$ is a distribution function characterizing the original non-turbulent plasma and introduce the following systematic expansion in the
turbulent and regular fields:
\begin{equation}
\delta f = \sum_{m=0}\sum_{n=0}\rho^{m}\tau^{n}\delta f_{mn}, \;\;\;  F^{\mu \nu} =  \sum_{m=0}\sum_{n=0}\rho^{m}\tau^{n}F_{mn}^{\mu\nu},
\end{equation}
where powers of $\rho$ count those of $F^R$ and powers of $\tau$ count those of $F^T$.
Turbulent polarization is described by contributions of the first order in the regular and the second in the turbulent fields.
The lowest nontrivial contribution to the induced current (\ref{incur}) is thus given by $\delta f_{12}$.  We have
\begin{equation}
\delta f  \simeq  \delta f_{\rm HTL} + \langle \delta f_{12} \rangle_{\rm I}  + \langle \delta f_{12} \rangle_{\rm II} \;  \nonumber
\end{equation}
where
\begin{eqnarray}
\delta f_{\rm HTL} & = & Gp_{\mu}F_{10}^{\mu\nu}\partial_{\mu,p}f^{\rm eq} \label{fhtl}  \nonumber \\
 \langle \delta f_{12} \rangle_{\rm I} & = & Gp_{\mu} \langle F_{01}^{\mu\nu}\partial_{\nu,p}Gp_{\mu^{\prime}}
 F_{10}^{\mu^{\prime}\nu^{\prime}}\partial_{\nu^{\prime},p}Gp_{\rho} F_{01}^{\rho
 \sigma}\rangle \partial_{\sigma,p} f^{\rm eq} \label{f12I} \nonumber \\
\langle \delta f_{12} \rangle_{\rm II} & = & Gp_{\mu} \langle  F_{01}^{\mu \nu}\partial_{\nu,p}Gp_{\mu^{\prime}}
F_{01}^{\mu^{\prime}\nu^{\prime}}\partial_{\nu^{\prime},p}Gp_{\rho} F_{10}^{\rho \sigma}
 \rangle \partial_{\sigma,p} f^{\rm eq} \label{f12II} \nonumber
\end{eqnarray}

\subsection{Nonlinear Landau damping and instabilities}

\begin{figure}[hbt]
\centering
\hspace{0.03\textwidth}
\includegraphics[height=0.25\textheight,width=0.45\textwidth]{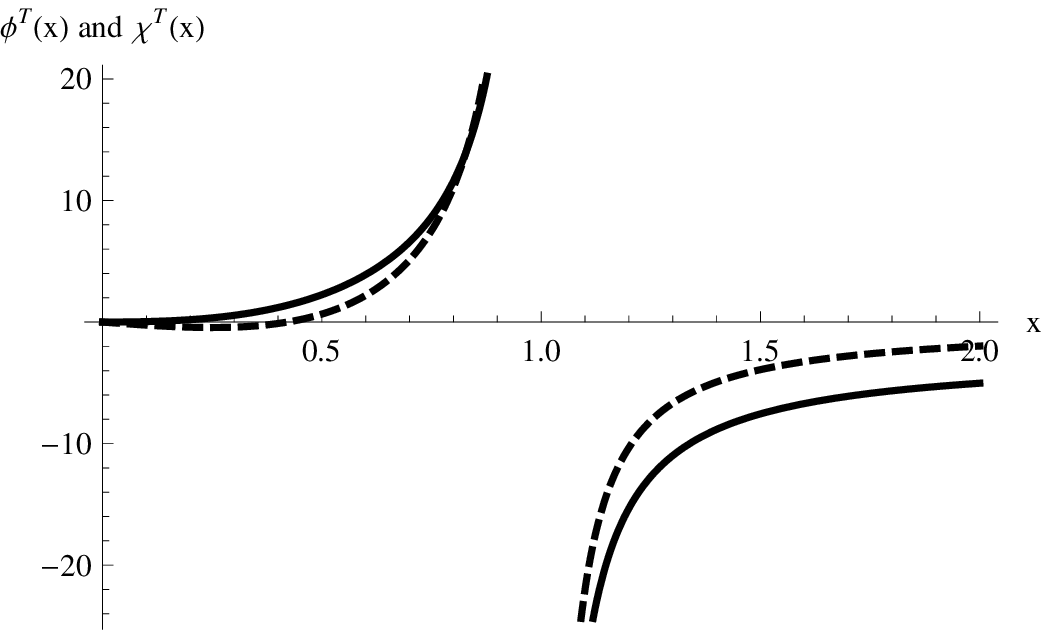}
\hspace{0.03\textwidth}
\includegraphics[height=0.25\textheight,width=0.45\textwidth]{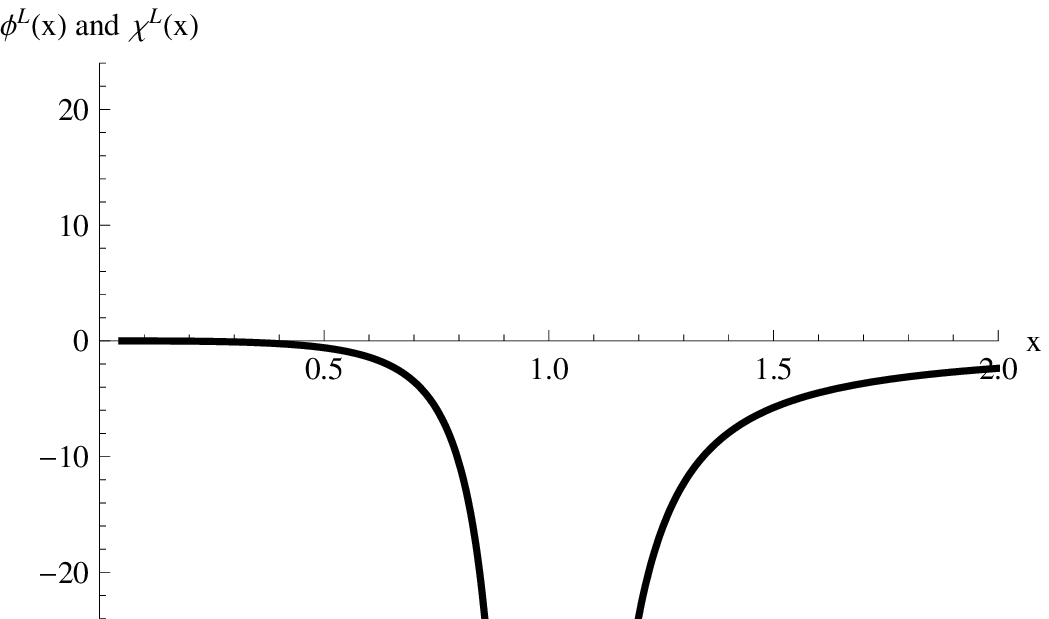}
\hspace{0.03\textwidth}
\caption{The polarization functions $(6\pi^{3/2}/e^{4})\, {\mathrm{Im}}\left[\phi^{\; (1)} (x)\right]$ (solid lines) and
$(6\pi^{3/2}/e^{4})\, {\mathrm{Im}}\left[\chi^{\; (1)} (x)\right]$ (dashed lines).
Left: transverse response; right: longitudinal response.}
\label{pct1}
\end{figure}

Generic expression for the polarization tensor taking into account turbulent effects can be written as
\begin{equation}
\Pi_{i j}(\omega,\mathbf k \, ; l)=\left(\delta_{i j}-\dfrac{k_{i}k_{j}}{k^{2}}\right)\Pi_{T}(\omega,\absvec{k}\, ; l)
+\dfrac{k_{i}k_{j}}{k^{2}}\Pi_{L} (\omega,\absvec{k} \, ; l) \label{poltendec}
\end{equation}
where $l \equiv \sqrt{2} (\tau a) / \sqrt{\tau^2+a^2}$. Both longitudinal and transverse components can be presented as a sum of Hard Thermal
Loops (HTL) contributions and the gradient expansion in the turbulent scale $l$:
\begin{eqnarray}
&& \Pi_{L(T)} (\omega,\mathbf{k} \, ; l)  =   \Pi^{\; \rm HTL}_{L(T)} (\omega,\mathbf{k}) +  \Pi^{\; \rm turb}_{L(T)} (\omega,\mathbf{k} \vert
\;
l) \label{poltenregturb} \\
&& \Pi^{\; \rm turb}_{L(T)} (\omega,\absvec{k} \, ; l)  =  \sum_{n=1}^\infty \dfrac{(\absvec{k}l)^n}{\mathbf{k}^2}
 \left [ \phi^{\; (n)}_ {L(T)} (x) \langle E_{\rm turb}^2 \rangle +
\chi^{\; (n)}_{L(T)} (x) \langle B_{\rm turb}^2 \rangle \right] \label{gradexp} \nonumber
\end{eqnarray}
$x=\omega/\absvec{k}$ and the standard HTL contribution
\begin{eqnarray}
&&\Pi_{L }^{\mathrm HTL} (\omega,\absvec{k}) =- m^2_D x^2 \left[1-\dfrac{x}{2} \; L(x) \right], \nonumber \\
&&\Pi_{T}^{\mathrm HTL} (\omega,\absvec{k})= m^2_D \dfrac{x^2}{2} \left[1+\dfrac{1}{2 x} \; (1-x^2 )\; L(x) \right] \nonumber \\
&& L(x) \equiv \ln\left|\dfrac{1+x}{1-x}\right|-\imath\pi\theta(1-x); \;\;\; m^2_D=e^2 T^2/3 .
\label{HTL}
\end{eqnarray}
The computation of turbulent polarization was carried out to second order in the gradient expansion \cite{KLM12}. In what follows we restrict
ourselves to discussing the leading contribution to the imaginary part of the polarization function corresponding to the turbulent modification
of Landau damping
in (\ref{HTL}):
\begin{eqnarray}
\mathrm{Im~} \Pi_{T} (\omega,\mathbf{k} \, ; l) &\simeq&- \pi m^2_D \dfrac{x}{4}(1-x^{2})\theta(1-x) \label{PIT} \\
&+& \dfrac{(\absvec{k} l)}{\mathbf{k}^{2}}\left(\left\langle E^{2}\right\rangle {\rm Im~} \phi^{(1)}_{\rm T} (x) +
\left\langle B^{2}\right\rangle {\rm Im~}\chi^{(1)}_{\rm T} (x)\right) \label{impartT} \nonumber \\
\mathrm{Im~} \Pi_{L} (\omega,\mathbf{k} \, ; l) &\simeq& -\pi m^2_D \dfrac{x^{3}}{2}\theta(1-x) \label{PIL} \\ & + & \dfrac{(\absvec{k}
l)}{\mathbf{k}^{2}} \left(\left\langle E^{2}\right\rangle {\rm Im~} \phi^{(1)}_{\rm L} (x) +\left\langle B^{2}\right\rangle {\rm Im~}
\chi^{(1)}_{\rm L} (x)\right), \label{impartL} \nonumber
\end{eqnarray}
The functions ${\rm Im~} \phi^{(1)}_{\rm L,T}$ and ${\rm Im~} \chi^{(1)}_{\rm L,T}$ are shown in Fig.~\ref{pct1}.
The conclusions following from Fig.~\ref{pct1} can be formulated as follows:
\begin{enumerate}
\item {\bf Timelike domain.} From Fig.~\ref{pct1} we see that the sign of the imaginary part of the turbulent
contribution to the polarization operator in the timelike domain $x>1$ is negative and corresponds to turbulent damping of timelike collective
excitations. This refers to both transverse and longitudinal modes. As the HTL contribution in this domain is absent, this turbulent damping is a
universal phenomenon present for all $\omega,k$ such that $\omega > k$ and all values of the parameters involved ($l$, $\langle B^2 \rangle$,
$\langle E^2 \rangle$). The turbulent damping leads to an attenuation of the propagation of collective excitations at some characteristic
distance.
\item {\bf Spacelike domain.} The situation in the spacelike domain $x<1$ is more diverse. In contrast with the timelike
domain the gradient expansion for the imaginary part of the polarization tensor starts from the negative HTL contribution corresponding to Landau
damping. As seen from Fig.~\ref{pct1} the imaginary parts of turbulent contributions to the longitudinal polarization tensor are
negative and are thus amplifying the Landau damping. The most interesting contributions come from the turbulent contributions to the transverse
polarization tensor. We see that the electric contribution ${\mathrm{Im}}[\phi_T^{\; (1)} (x)]$ in the spacelike domain is positive at
all $x$ while the magnetic contribution ${\mathrm{Im}}[\chi_T^{\; (1)} (x)]$ is negative for $x<x^* \approx 0.43$ and positive for $x>x^*$.
This means that the turbulent plasma becomes unstable for sufficiently strong turbulent fields.
\end{enumerate}

\section{Comments and conclusions}\label{conc}

Let us briefly discuss some relevant points:

\medskip

\noindent {\small \bf 1.} The above presented results are obtained in the framework of a perturbative expansion based on two crucial assumptions.
First, one assumes  slow temporal evolution of the distribution function due to particle interaction with turbulent fields thus neglecting the
corresponding $f_{0n}$ contributions. Second, changes in the distribution function are treated as small. This, in turn, means that turbulent
fields should be small enough. In this sense the reliable results refer to small modifications of Landau damping but, as the onset of turbulent
instability takes place for parametrically large fields, this result should be considered as a qualitative indication.

\noindent {\small \bf 2.} The observed instability can be termed ``secondary'' because the turbulent fields themselves result from some ``first
level'' instabilities. The origin of the effect is in turbulent stochastic inhomogeneity and thus similar to stochastic transition radiation,
which vanishes in the limit $l \to 0$) \cite{T72}.

\noindent {\small \bf 3.} The non-Abelian generalization of the above-described results for the imaginary part of the polarization tensor leads to
identical expressions, the only difference being in trivial color factors, just as in the HTL case.

\end{document}